\documentclass[fleqn]{2023SCGE}
\begin{document}
\ensubject{subject}
\ArticleType{Article}
\SpecialTopic{SPECIAL TOPIC: }
\Year{2024}
\Month{February}
\Vol{xxxx}
\No{xxxx}
\DOI{xx}
\ArtNo{xxxx}
\ReceiveDate{xxxx}
\AcceptDate{xxxx}
\title{Investigating the neutron star physics through observations of several young pulsars in the dipole-field re-emergence scenario}{Investigating the neutron star physics through observations of several young pulsars in the dipole-field re-emergence scenario}

\author[1]{Yu-Long Yan}{}%
\author[1]{Quan Cheng}{qcheng@ccnu.edu.cn}
\author[1,2]{Xiao-Ping Zheng}{}

\AuthorMark{Yu-Long Yan}
\AuthorCitation{Yu-Long Yan, Quan Cheng and Xiao-Ping Zheng}

\address[1]{Institute of Astrophysics, Central China Normal University, Wuhan 430079, China}
\address[2]{Department of Astronomy, Huazhong University of Science and Technology, Wuhan 430074, China}



\abstract{The observed timing data, magnetic tilt angle $\chi$, and age of young pulsars could be used to probe some important issues about neutron star (NS) physics, e.g., the NS internal magnetic field configuration, and the number of precession cycles $\xi$. \textbf{Both} quantities are critical in studying the continuous gravitational wave emission from pulsars, and the latter generally characterizes the mutual interactions between superfluid neutrons and other particles in the NS interior. The timing behavior of pulsars can be influenced by the dipole field evolution, which instead of decaying, may increase with time. An increase in the dipole field may result from the re-emergence of the initial dipole field $B_{\rm d,i}$ that was buried into the NS interior shortly after the birth of the NS. In this work, the field re-emergence scenario $\xi$ and the internal field configuration of several young pulsars, as well as their $B_{\rm d, i}$ are investigated by assuming typical accreted masses $\Delta M$. Moreover, since the Crab pulsar has an exactly known age and its tilt angle change rate can be inferred from observations, we can set stringent constraints on its $\xi$, $B_{\rm d,i}$, and $\Delta M$. Although for other young pulsars without exactly known ages and tilt angle change rates, these quantities cannot be accurately determined, we find that their $\xi$ are generally within $\sim10^4-10^6$, and some of them probably have magnetar-strength $B_{\rm d,i}$. Our work could be important for investigating the transient emissions associated with NSs, the origin of strong magnetic fields of NSs, pulsar population, continuous gravitational wave emission from pulsars, and accretion under extreme conditions in principle.}

\keywords{pulsar, neutron star, magnetic field}

\PACS{97.60.Gb, 97.60.Jd, 95.85.Sz}

\maketitle

\begin{multicols}{2}
\section{Introduction}\label{sec:intro}
Neutron stars (NSs) are a class of compact objects that possess extremely high densities, strong magnetic fields, and strong gravitational fields in the Universe \cite{Shapiro:1983}. Knowledge of the physical properties of NSs, especially the magnetic fields and the neutron superfluid physics at supranuclear densities are vital for understanding a variety of astrophysical phenomena, such as gamma-ray bursts \cite{Dai:1998,Zhang:2001,Metzger:2008}, superluminous supernovae \cite{Kasen:2010}, fast blue optical transients\cite{Yu:2015,Liu:2022},
\Authorfootnote
fast radio bursts 
\cite{Zhang:2014,Dai:2016,CHIME:2020},
soft gamma-ray repeaters \cite{Duncan:1992,Olausen:2014}, anomalous X-ray pulsars \cite{Thompson:1996}, central compact objects 
\cite{Gotthelf:2005,Page:2011}, and pulsar glitches \cite{Andersson:2012,Pizzochero:2017}. Currently, there are some open issues about the NS magnetic fields. The first one is the strength of the initial dipole field of NSs, though it is proposed that different types of NSs (magnetars, high-magnetic-field radio pulsars, ordinary pulsars, and central compact objects) may have different initial dipole fields \cite{Viganò:2013,Dirson:2022}. This issue is quite critical since the initial dipole fields are a key parameter in accounting for the light curves of some transient emissions in the NS-powered scenario \cite{Yu:2007,Lü:2015,Xie:2022,Inserra:2013,Yu:2017,Yu:2015,Liu:2022} and the origin of strong magnetic fields of NSs \cite{Blandford:1983,Duncan:1992,Ferrario:2006,Makarenko:2021,Mösta:2015,Raynaud:2020}, and the study of pulsar population \cite{Faucher:2006,Popov:2010,Gullón:2014,Dirson:2022}

The second issue is about the internal field configuration of NSs. Though the results of numerical simulations suggested that a twisted-torus shape consisting of both poloidal and toroidal fields may exist in the NS interior \cite{Braithwaite:2004}, the dominant component of the internal fields is still under debate (see, e.g., Refs\cite{Ciolfi:2010,Mastrano:2011,Dall’Osso:2015}). With the presence of strong internal fields, the NSs will deform into a quadrupole ellipsoid.\footnote{In this paper, we only consider the magnetic deformation of NSs and neglect the elastic deformation of NS crusts. Our arguments are as follows. Though the maximum ellipticity that the solid NS crust can sustain maybe $\sim10^{-7}$, the actual ellipticity of the crust is still unknown, which depends on the mechanisms that drive the deformation \cite{Jones:2024}. In the case of centrifugal-force driven deformation, the ellipticity is estimated to be $\sim10^{-11}$ for the Crab pulsar \cite{Cutler:2002,Dall'Osso:2017}, which is much smaller than the amplitude of its magnetically-induced ellipticity $\sim10^{-8}$ (see Sec. \ref{sec:model}). For other young pulsars with stronger magnetic fields but slower spins, their magnetically-induced ellipticities would be even larger than the ellipticities of elastic deformation of the solid crusts.} The sign of the ellipticity $\epsilon_{\rm B}$ of the deformed NS is determined by the NS internal field configuration \cite{Bonazzola:1996,Cutler:2002}. In the poloidal-toroidal twisted-torus scenario, the poloidal-dominated (PD) and toroidal-dominated (TD) configurations respectively correspond to $\epsilon_{\rm B}>0$ and $\epsilon_{\rm B}<0$ \cite{Ciolfi:2010,Dall’Osso:2015}. Generally, the deformed NS is not in a state with minimum spin energy because its spin and magnetic axes are neither aligned nor orthogonal with each other in most cases \cite{Dall’Osso:2009,Biryukov:2017,Nikitina:2017}. To achieve this state,  precession of the NS's magnetic axis around the spin axis will take place, resulting in the evolution of magnetic tilt angle between the two axes \cite{Dall’Osso:2009}. Precession of the deformed NS can be used to account for various of observations, such as the periodic variations of the pulse profiles of PSR B1828-11 \cite{Stairs:2000} and the timing residuals of PSR B1642-03 \cite{Shabanova:2001}, the possible existence of periodic modulations in the X-ray afterglows of gamma-ray bursts \cite{Suvorov:2020}, the periodic modulations in the pulsed hard X-ray emission of the magnetars 4U 0142$+$61, 1E 1547--5408, and SGR 1900$+$14 \cite{Makishima:2014,Makishima:2021a,Makishima:2021b}, the periodicities detected in the repeating fast radio bursts (FRBs) 180916.J0158$+$65 and 121102 \cite{Levin:2020,Wasserman:2022}. Moreover, the precession of the NS may also leave imprints in its continuous gravitational wave (GW) emission, which might be discovered by the next-generation ground-based GW detectors \cite{Gao:2020}. The GW emission from magnetic deformation of the NS is maximized when the star has a magnetic tilt angle $\chi=\pi/2$, which can be realized through precession of the NS if its internal fields are TD ($\epsilon_{\rm B}<0$) \cite{Cutler:2002,Stella:2005,Cheng:2019}. On the contrary, if the NS has PD ($\epsilon_{\rm B}>0$) internal fields, its tilt angle will gradually decrease and thus the GW emission will gradually weaken with the precession of the NS \cite{Dall’Osso:2009,Cheng:2019}. Since the precession behavior and tilt angle evolution of the NS are dependent on the NS's internal field configuration, and different configurations may result in diverse features in both electromagnetic and gravitational emission, a detailed investigation on the NS's internal field configuration is necessary.

Actually, during the precession process of the NS, besides the internal field configuration, the number of precession cycles, $\xi=\tau_{\rm dis}/P_{\rm pre}$, is also a key parameter that can affect the tilt angle evolution \cite{Cutler:2002,Dall’Osso:2009,Cheng:2019}, where $\tau_{\rm dis}$ and $P_{\rm pre}$ are respectively the dissipation time scale of the precessional energy and precession period of the NS. The quantity $\xi$ represents certain viscous mechanisms through which the NS's precessional energy is dissipated \cite{Cutler:2002,Dall’Osso:2009}, and determines the tilt angle change rate due to viscous dissipation of the precessional energy \cite{Cutler:2002,Dall’Osso:2009,Cheng:2019}. More importantly, its value is approximately equal to the reciprocal of the superfluid mutual friction parameter, ${\mathcal{B}}$, and thus can be associated with complicated mutual interactions between superfluid neutrons and other particles in the NS interior \cite{Haskell:2018,Cheng:2019,Hu:2023a}. Though theoretical work has shown that the superfluid mutual friction parameter may distribute in a wide range $10^{-8}\lesssim{\mathcal{B}}\lesssim10^{-1}$ (which corresponds to $10 \lesssim\xi\lesssim 10^8$) \cite{Haskell and Sedrakian:2018}, and more stringent constraints on ${\mathcal{B}}$ have been obtained by modeling the rise processes of some large glitches of the Crab and Vela pulsars \cite{Haskell:2018,Ashton:2019}, the specific values of $\xi$ are still under debate. An investigation on $\xi$ is not only important for understanding of the mutual interactions between superfluid neutrons and other particles but also crucial for the study of continuous GW emission and timing behaviors of pulsars since $\xi$ can remarkably affect the tilt angle evolution \cite{Stella:2005,Cheng:2019,Hu:2023a}. As a feasible way, by combining the observed timing data and tilt angles of young pulsars with the theory of dipole field decay, $\xi$ were constrained in previous work, and the results for the Crab and Vela pulsars are generally consistent with that obtained from glitch observations \cite{Cheng:2019,Hu:2023a,Yan:2024}. However, the dipole fields of young pulsars may not merely manifest as decay, especially when considering that these pulsars are relatively young. In fact, rather than decay with time, their dipole fields may increase gradually at current ages.

Depending on the explosion energies of supernovae, internal structures of progenitor stars, initial dipole magnetic fields and spin periods of proto-neutron stars, and kick velocities gained by the young pulsars \cite{Geppert:1999,Fryer:2006}, fallback accretion may happen in young pulsars. Some indirect observational evidence indeed indicates that the Vela pulsar could be surrounded by an active fallback disk \cite{Ozsukan:2014}. It should be stressed that we mainly focus on the effect of accreted matter on the dipole field evolution of NSs, whether the fall-back accretion is Bondi-like or disk-like \cite{Geppert:1999} is beyond the scope of this work. If the young pulsars experienced a fall-back accretion phase shortly after their births, it is possible that their initial dipole fields were submerged into the NS interior by the accreted matter and then diffuse out to the stellar surface gradually due to Ohmic diffusion, leading to the increase of the dipole fields \cite{Muslimov:1995,Geppert:1999,Shabaltas:2012,Fu:2013,Ho:2015,Igoshev:2016,Cheng:2023}. Moreover, re-diffusion of the magnetic fields that are buried into the crust in some cases could heat star \cite{Gourgouliatos:2020}. Since the braking indices of pulsars could be affected by the re-emergence of the submerged dipole fields, revisiting the constraints on the internal field configuration and $\xi$ of young pulsars in the field re-emergence scenario is necessary. This is the main purpose of our work and also represents the main difference in comparison with previous work \cite{Cheng:2019,Hu:2023a,Yan:2024}. Our results show that in the dipole-field re-emergence scenario, the internal field configurations and the values of $\xi$ of young pulsars cannot be determined from the braking index and tilt angle observations as proposed in previous work for the dipole-field decay scenario. In fact, the mass accreted onto the NSs during the fall-back accretion process can also affect the final results. Assuming different accreted masses, we obtain the ranges of both the initial dipole fields and $\xi$ of the young pulsars. For the Crab pulsar, by using the dipole field increase rates derived in previous work \cite{Yan:2024}, we further set stringent constraints on its initial dipole field, accreted mass, and $\xi$. The results are important for the study of internal field configurations, superfluid mutual frictions, and initial dipole fields of young pulsars. Moreover, our results may also shed light on accretion physics under extreme conditions because the accreted mass of the Crab is also obtained.

This paper is organized as follows. In Sec. \ref{sec:model}, we show the theoretical model and observational data of the young pulsars used. Constraints on the internal field configurations, initial dipole fields, and $\xi$ of these pulsars are presented in Sec. \ref{sec:results}. Based on the inferred tilt angle change rate of the Crab pulsar \cite{Lyne:2013}, we further constrain the initial dipole field, accreted mass, and $\xi$ of the Crab pulsar in this section. Finally, conclusions and discussions are given in Sec. \ref{sec:conclusion}.

\section{Theoretical model and observational data of young pulsars}\label{sec:model}
Generally, a magnetically deformed NS embedded in a plasma-filled magnetosphere \cite{Goldreich:1969,Spitkovsky:2006} could spin down because of magnetospheric currents (MCs) and GW radiation \cite{Shapiro:1983}. The spin-down evolution of such a NS can be described by the following formula \cite{Cheng:2019}
\begin{equation}
\begin{aligned}
    \dot{\omega}=-\frac{B_{\mathrm{d}}^{2} R^{6} \omega^{3}}{6 I c^{3}} (1+\sin ^{2} \chi)-\frac{2 G \epsilon_{\mathrm{B}}^{2} I \omega^{5}}{5 c^{5}} \sin ^{2} \chi\left(1+15 \sin ^{2} \chi\right),
\end{aligned}
\label{dwdt}
\end{equation}
where $\omega$, $B_{\rm d}$, $R$, $I$, and $\chi$ are respectively the NS's angular frequency, dipole magnetic field, radius, moment of inertia, and magnetic tilt angle. The ellipticity of magnetic deformation $\epsilon_{\rm B}$ depends on the NS equation of state, internal magnetic energy, and internal magnetic field configuration (e.g., \cite{Haskell:2008,Mastrano:2011}). As mentioned in Sec. \ref{sec:intro}, different internal field configurations correspond to different signs of $\epsilon_{\rm B}$. Following Cheng et al. \cite{Cheng:2019}, we adopt two possible forms of $\epsilon_{\rm B}$ that were obtained with the effect of proton superconductivity being taken into account and correspond to the PD and TD internal fields, respectively. The specific forms are $\epsilon_{\rm B}=3.4\times10^{-7}(B_{\rm d,i}/10^{13}~{\rm G})(H_{\rm c1}(0)/10^{16}~{\rm G})$ for the PD case, and $\epsilon_{\rm B}\simeq-10^{-8}(\bar{B}_{\rm in}/10^{13}~{\rm G})(H/10^{15}~{\rm G})$ for the TD case, where $B_{\rm d,i}$ is the initial strength of the dipole field, and $H_{\rm c1}(0)=10^{16}$ G and $H\simeq10^{15}$ G are respectively the critical field strengths of the two cases \cite{Akgün:2008,Lander:2013,Cheng:2019}. $\bar{B}_{\rm in}$ in the above formula is the volume-averaged strength of the internal toroidal field. A detailed discussion about the two forms of $\epsilon_{\rm B}$ used here can be found in \cite{Cheng:2019}. It should be pointed out that the precession of the NS is actually not free because of the existence of surface dipole field. In this case, two torques, namely the far-zone and near-zone radiation torques can act on the NS and may result in periodic variations in the NS's spin-down rate $\dot{\omega}$ \cite{Melatos:1999,Link:2001,Zanazzi:2015}. However, since the stellar \textbf{quadrupole} deformation is mainly caused by the internal fields, the effects of the two radiation torques on $\dot{\omega}$ could be neglected.  

Defining the ratio of MCs to GW spin-down rates $\eta=5c^2B_{\rm d}^2R^6\left(1+\sin^2\chi\right)/\left[12G\epsilon_{\rm B}^2I^2\omega^2\left(1+15\sin^2\chi\right)\sin^2\chi\right]$, we have $\eta\gg1$ because for young pulsars the GW spin-down is generally much smaller than the MCs spin-down \cite{Cheng:2019}. Therefore, after neglecting the GW spin-down in Eq. (\ref{dwdt}), the dipole field $B_{\rm d}$ can be expressed as a function of $\chi$ as 
\begin{equation}
\begin{aligned}
B_{\rm
d}=\left[-\frac{6\dot{\omega}Ic^3}{\omega^3R^6(1+{\rm
sin}^2\chi)}\right]^{1/2}.
\end{aligned}
\label{Bdx}
\end{equation}
Eq. (\ref{Bdx}) suggests that by using the observed spin period $P$, its fisrt time derivative $\dot{P}$, and $\chi$ of a pulsar, its $B_{\rm d}$ can be estimated since we have $\omega=2\pi/P$ and $\dot{\omega}=-2\pi\dot{P}/P^2$ \cite{Cheng:2019,Hu:2023a}. The same as in previous work \cite{Cheng:2019,Hu:2023a,Yan:2024}, here we take $I=10^{45}\, \rm{g\, cm^2}$ and $R=10$ km.  

With the spin-down of the NS, both the electromagnetic (EM) radiation in the magnetosphere and GW radiation can lead to aligned torques between the magnetic and spin axes \cite{Cutler:2000,Dall’Osso:2009,Philippov:2014}, resulting in the decrease of the tilt angle $\chi$.\footnote{Though it was proposed that the magnetosphere effect may lead to the increase of $\chi$ \cite{Beskin:1993,Arzamasskiy:2017}, numerical simulations are still needed to verify the result.} Meanwhile, damping of the stellar precession due to internal viscosity \cite{Dall’Osso:2009,Cheng:2019} can either increase or decrease $\chi$, which relies on the sign of $\epsilon_{\rm B}$ \cite{Cutler:2002,Dall’Osso:2009,Cheng:2019}. The change rate of $\chi$ due to GW radiation is given by Eq. (2.9) in \cite{Cutler:2000}. Following Ref. \cite{Philippov:2014}, the contribution from EM radiation can be obtained by substituting their Eqs. (9) and (15) into Eq. (5). Here we take the coefficient of this term to be 1/6 \cite{Cheng:2019} as shown in Eq. (\ref{chidot}) below, instead of 1/4 in \cite{Philippov:2014}. This has little effect on our final results. The evolution of $\chi$ caused by damping of the NS's precession due to internal viscosity is $\dot{\chi}=\cot\chi/\tau_{\rm dis}$ for $\epsilon_{\rm B}<0$ (see Eq. (14) in \cite{Dall’Osso:2009}), where the dissipation time scale of precession is $\tau_{\rm dis}=\xi P/\left|\epsilon_{\rm B}\right|$ \cite{Cutler:2002}. Thus the angle evolution contributed by internal dissipation of the precession is $\dot{\chi}=-\epsilon_{\rm B}\cot\chi/\xi P$ for $\epsilon_{\rm B}<0$. Similarly, using the first part of Eq. (A8) in \cite{Dall’Osso:2009}, we can derive $\dot{\chi}=-\epsilon_{\rm B}\tan\chi/\xi P$ for $\epsilon_{\rm B}<0$. To sum up, for the magnetically deformed rotating NS with a plasma-filled magnetosphere,\footnote{Although Faucher-Giguère \& Kaspi \cite{Faucher:2006} suggested that the effective magnetic field and tilt angle of a NS may not decay over a timescale of $\sim10^8$ yrs, the inclusion of the aligned torque due to EM radiation in the tilt angle evolution can make the results of population synthesis studies be more consistent with pulsar observations \cite{Philippov:2014,Gullón:2014,Dirson:2022}. However, it is still unknown whether the results of population synthesis could be consistent with pulsar observations after involving the aligned torques from EM and GW radiation, and the effect of viscous damping of the stellar precession. This needs to be verified in future studies.} the tilt angle change rate can be expressed as \cite{Cutler:2000,Jones:2001,Philippov:2014,Cheng:2019}
\begin{eqnarray}
\dot{\chi}=\left\{ \begin{aligned}
         -\frac{2G}{5c^5}I\epsilon_{\rm
B}^2\omega^4&\sin\chi\cos\chi(15\sin^2\chi+1)-{\epsilon_{\rm B}\over\xi P}{\rm tan}\chi\\&-\frac{B_{\rm d}^2R^6\omega^2}{6Ic^3}\sin\chi\cos\chi,~{\rm for}~\epsilon_{\rm B}>0 \\
                  -\frac{2G}{5c^5}I\epsilon_{\rm
B}^2\omega^4&\sin\chi\cos\chi(15\sin^2\chi+1)-{\epsilon_{\rm
B}\over\xi P}{\rm cot}\chi\\&-\frac{B_{\rm
d}^2R^6\omega^2}{6Ic^3}\sin\chi\cos\chi,~{\rm for}~\epsilon_{\rm
B}<0.
                          \end{aligned} \right.
\label{chidot}
\end{eqnarray}
We point out that the first term in Eq. (\ref{chidot}) can be neglected in principle as the alignment effect due to GW radiation is quite weak. However, this term is still retained for completeness.

During the spin-down of the NS, its dipole field may not remain constant. In previous work \cite{Cheng:2019,Hu:2023a,Yan:2024}, we mainly focus on the decay of the dipole field owing to the non-zero resistance of the NS crust and relative motions between the electrons and neutrons in the core. The dipole field, in this case, could decay due to Hall drift and Ohmic dissipation if it originates from the crust, or decay due to ambipolar diffusion provided that it has a core origin \cite{Goldreich:1992,Aguilera:2008,Passamonti:2017,Kojima:2020,Skiathas:2024}. However, it is also possible that with the spin-down of the NS, its dipole field may increase with time. The increase of the dipole field may be caused by the re-emergence of the initially higher dipole field, which was submerged into the NS interior during the early fall-back accretion phase after the birth of the NS \cite{Geppert:1999,Ho:2015,Cheng:2023,Bernal:2013,Zhong:2021}. The field re-diffusion rate $\dot{B}_{\rm d}$ is determined by the specific mechanisms that act in the re-emergence process. To be specific, $\dot{B}_{\rm d}$ can be calculated by solving the MHD induction equation (see, e.g., Refs\cite{Geppert:1999,Ho:2011,Fu:2013}) after assuming that the buried dipole field gradually diffuses out to the NS surface because of Ohmic diffusion. Since the Ohmic diffusion time scale depends on the electrical conductivity and temperature profile in the NS interior, in the calculations of \cite{Ho:2011}, detailed NS physics, such as new results on the electrical and thermal conductivities, and dense matter equation of state were all involved. Based on the numerical results of field re-diffusion performed in Refs\cite{Geppert:1999,Ho:2011}, a simple analytic expression for $\dot{B}_{\rm d}$ was obtained by Fu et al. \cite{Fu:2013}, which has the following form
\begin{equation}\label{Bdot}
    \dot{B}_{\rm d}=10^{-11}\left(\frac{\Delta M}{M_{\odot}}\right)^{-3}\left(1-\frac{B_{\rm d}}{B_{\rm d,i}}\right)^{2}~{\rm G/yr}.
\end{equation}
In the above formula, $\Delta M$ is the mass of the accreted matter. Denoting $B_{\rm d,r}$ to be the residual strength of the dipole field at the time when submergence terminated, its current strength is 
\begin{equation}\label{Bd}
    B_{\rm d}=B_{\rm d,r}+\int_0^t  \dot{B}_{\rm d}dt,
\end{equation}
where $t$ is the pulsar's age. For the young pulsars whose ages can be inferred from the associated supernova remnants, $t$ is taken to be the ages of supernova remnants (see Ref. \cite{Liu:2024}), though for most sources, only a rough range is given \cite{Liu:2024}. Two exceptions, namely the Crab pulsar and PSR J1734-3333, should be paid attention to. The former has an exactly known age $t=970$ yr, while the supernova-remnant age of the latter is unavailable currently. Thus, for the latter, $t$ is adopted to be its characteristic age. The ages of the young pulsars are shown in Tab. \ref{tab:table}.

In this work, the dipole-field evolution is divided into two stages for simplicity. The first stage is the submergence stage, in which only the decrease of the dipole field due to burial is considered. The second one is the re-emergence stage, in which only the increase of dipole field due to re-emergence is taken into account. In other words, the field decay because of Hall drift and Ohmic dissipation are neglected in both stages. For the young pulsars considered in this work (see Tab. \ref{tab:table}), they are possibly in the second stage since direct observational evidence for the existence of a fall-back disk surrounding these pulsars is still lacking. Moreover, the fall-back accretion phase during which the dipole field is buried may only last for months \cite{Chevalier:1989,Geppert:1999}, much shorter than the ages of the pulsars. The residual strength of the dipole field at the end of the first stage can be given by \cite{Shibazaki:1989}
\begin{equation}\label{Bdr}
    B_{\rm d,r}=\frac{B_{\rm d,i}}{1+\Delta M / 10^{-5} M_{\odot}}.
\end{equation}
For typical accreted mass $\Delta M=10^{-5}-\-10^{-3}M_{\odot}$ \cite{Shibazaki:1989}, the residual strength will be $0.5-\-0.01$ times of the initial strength.

After taking into account the spin, tilt angle, and dipole field evolutions, the braking index of the NS has the following form
\begin{equation}\label{bi}
\begin{aligned}
n=& 3-\frac{2 P}{\dot{P}}\left\{\frac{\dot{B}_{\mathrm{d}}}{B_{\mathrm{d}}}+\dot{\chi} \sin \chi \cos \chi\left[\frac{1}{1+\sin ^{2} \chi}\right.\right.\\
&\left.\left.+\frac{1+30 \sin ^{2} \chi}{\eta \sin ^{2} \chi\left(1+15 \sin ^{2} \chi\right)}\right]\right\}.
\end{aligned}
\end{equation}

\begin{table*}[t]
\caption{\label{tab:table}%
The age $t$, spin period $P$, first derivative $\dot{P}$, and measured magnetic tilt angle $\chi$ of young pulsars with a steady braking index $n$. The * indicates that the characteristic age is adopted.
}
\begin{tabular}{ccccccc}
\toprule
Pulsar& $t$ (kyr)& $P$ (s)& $\dot{P}~(10^{-13}~\rm{s/s})$& $n$& $\chi$& References
\\ \hline
PSR J1734-3333 & 8.13* & 1.17 & 22.8 & $0.9 \pm\, 0.2$ & $6^{\circ}$, $21^{\circ}$ & \cite{Nikitina:2017,Espinoza:2011} \\
PSR B0833-45 (Vela) & 9-27 & 0.089  & 1.25 & $1.4 \pm\, 0.2$ & $62^{\circ}$,$70^{\circ}$,$75^{\circ}$,$79^{\circ}$ &  \cite{Liu:2024,Lyne:1996,Dyks:2003,Watters:2009,Barnard:2016} \\
PSR J1833-1034 & 1.55-1.8 & 0.062 & 2.02 & $1.8569 \pm\, 0.0006$ & $70^{\circ}$ & \cite{Liu:2024,Roy:2012,Li:2013} \\
PSR J1846-0258 & 1.69-1.85 & 0.324 & 71 & $2.19 \pm\, 0.03$ & $10^{\circ}$ & \cite{Liu:2024,Archibald:2015,Wang:2014} \\
PSR B0531+21 (Crab) & 0.97 & 0.033 & 4.21 & $2.51 \pm\, 0.01$ & $45^{\circ}$,$60^{\circ}$,$70^{\circ}$ & \cite{Dyks:2003,Watters:2009,Lyne:1993,Harding:2008,Du:2012} \\
PSR J1119-6127 & 4.2-7.1 & 0.408 & 40.2 & $2.684 \pm\, 0.002$ & $7^{\circ}$,$16^{\circ}$,$21^{\circ}$ & \cite{Liu:2024,Weltevrede:2011,Rookyard:2015a,Rookyard:2015b,Tian:2018} \\
PSR J1513-5908 & 1.9-1.9 & 0.151 & 15.3 & $2.839 \pm\, 0.001$ & $3^{\circ}$,$10^{\circ}$ &  \cite{Liu:2024,Nikitina:2017,Livingstone:2007}\\
PSR J1640-4631 & 1-8 & 0.207 & 9.72 & $3.15 \pm\, 0.03$ & $-\-$ & \cite{Liu:2024,Archibald:2016} \\
\bottomrule
\end{tabular}
\end{table*}
 
\section{Results}\label{sec:results} 

Same as in previous work \cite{Hu:2023a,Yan:2024}, here we only focus on the young pulsars with a steady $n$, and the error bars in $n$ are neglected in the calculations. The ages, measured timing data, and tilt angles of these pulsars are listed in Tab. \ref{tab:table}. Our calculations are performed as follows. Substituting Eqs. (\ref{Bdot}) and (\ref{Bdr}) into Eq. (\ref{Bd}), one can find that there are two variables, i.e., $\Delta M$ and $B_{\rm d,i}$ in this equation because $B_{\rm d}$ can be obtained from Eq. (\ref{Bdx}) after taking a value for $\chi$ and the values of $t$ can be found in Tab. \ref{tab:table}. Therefore, taking a value for $\Delta M$, we can solve for $B_{\rm d,i}$ through Eq. (\ref{Bd}), and then obtain $\dot{B}_{\rm d}$ through Eq. (\ref{Bdot}). Obviously, the vaules of $B_{\rm d,i}$ and $\dot{B}_{\rm d}$ depend on the specific values of $\chi$, $t$, and $\Delta M$ adopted. Substituting the calculated $B_{\rm d}$, $\dot{B}_{\rm d}$, Eq. (\ref{chidot}), and the timing data of a pulsar into Eq. (\ref{bi}), we can solve for the curve of $\xi$ versus $\chi$ for given values of $\Delta M$ and $t$. We stress that an appropriate expression for $\dot{\chi}$ as shown in Eq. (\ref{chidot}), should be adopted when solving Eq. (\ref{bi}). \textbf{Unless otherwise} specified, the internal toroidal field is assumed to be ten times the initial surface dipole field in the TD case, i.e., $\bar{B}_{\rm in}=10B_{\rm d,i}$ \cite{Cheng:2019,Hu:2023a,Yan:2024} as the internal toroidal field was possibly not affected by the submergence and re-emergence of the surface dipole field. It should be pointed out that to derive the $\xi-\-\chi$ curve, it is actually unnecessary to solve the evolutionary equations of $B_{\rm d}$, $\chi$, and $\omega$ simultaneously. The arguments are as follows. First, the initial values for $\chi$ and $\omega$ of a pulsar are also required when solving these equations, which are hard to determine. Second, the values of $P$, $\dot{P}$, $n$, $\chi$, $B_{\rm d}$, and $\dot{B}_{\rm d}$ used in Eq. (\ref{bi}) are all current values of a pulsar, the results obtained are current snapshots of the $\xi-\-\chi$ relation during the whole evolution of the pulsar.

\begin{figure}[H]
\centering
\includegraphics[scale=0.5]{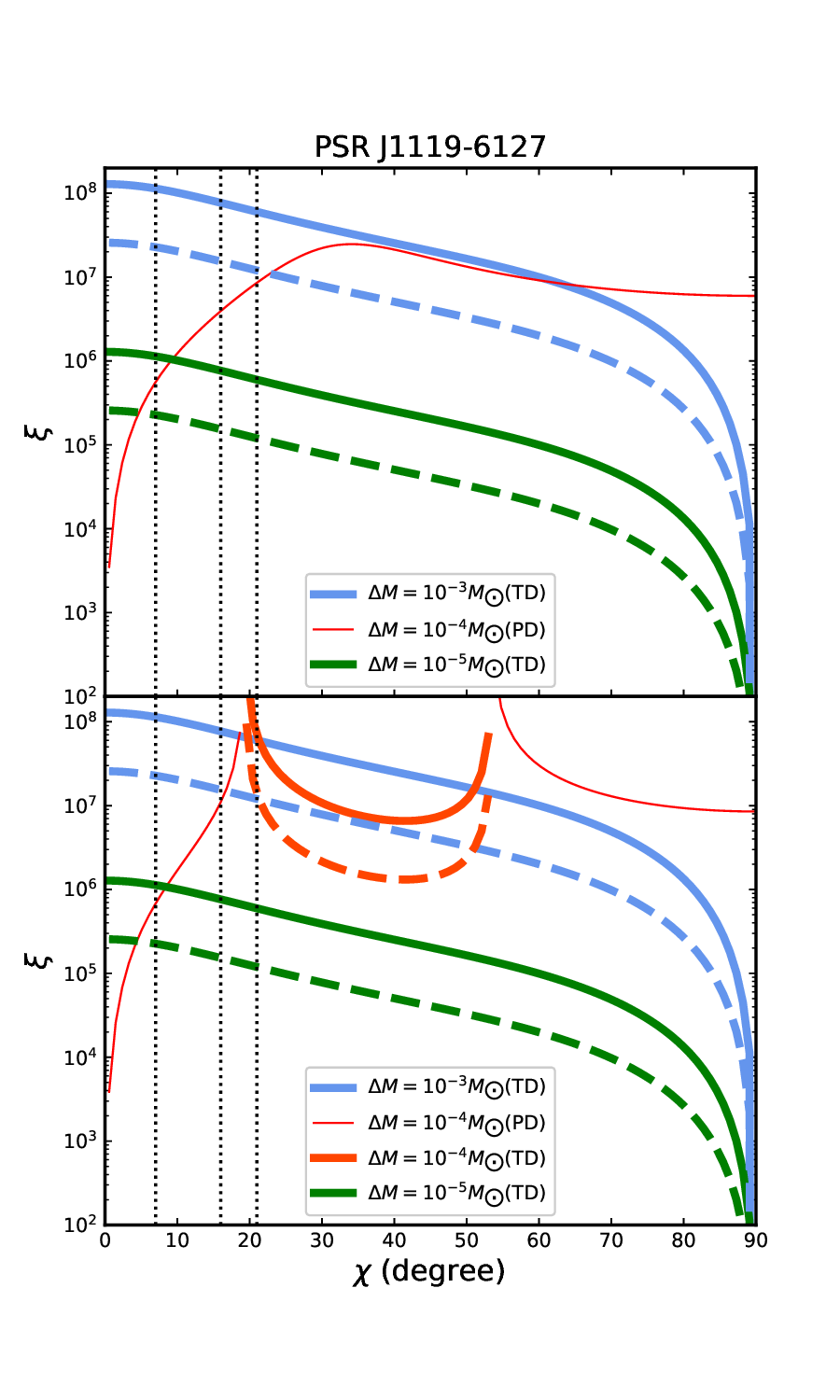}
\caption{The number of precession cycles $\xi$ versus tilt angle $\chi$ for PSR J1119-6127. See the text for details.}
\label{fig:1}
\end{figure}

\begin{figure}[H]
\centering
\includegraphics[scale=0.5]{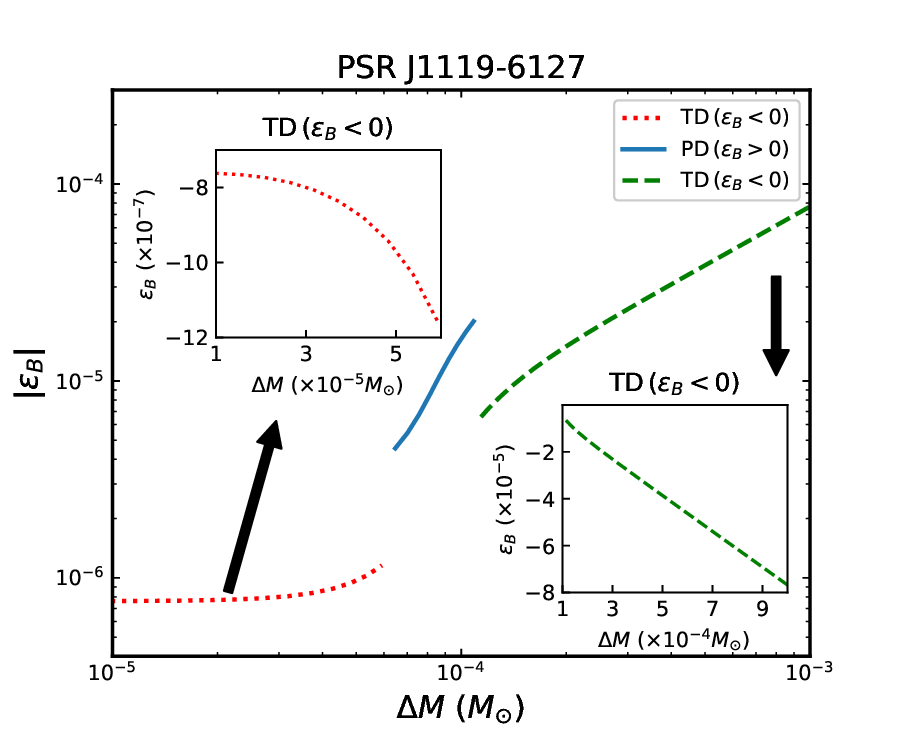}
\caption{The absolute value of ellipticity $\epsilon_{B}$ versus accreted mass $\Delta M$ for PSR J1119-6127. Here, its age and tilt angle are respectively taken as $t=4.2$ kyr and $\chi=21^\circ$. See the text for details.}
\label{fig:2}
\end{figure}

We take PSR J1119-6127 as an example and give a detailed explanation of its results since diagnosing the internal field configuration of this pulsar is relatively complex. The curves of $\xi$ versus $\chi$ at three typical accreted masses $\Delta M$ (see the legends) for PSR J1119-6127 are presented in Fig. \ref{fig:1}. The labels "TD" and "PD" in the legends indicate that in this case, the pulsar has TD and PD internal fields, respectively. The upper and lower panels respectively show the results of $t=4.2$ and 7.1 kyr \cite{Liu:2024}. The thin solid line in the upper panel indicates that the pulsar has PD internal fields if it has an age of $4.2$ kyr and accreted a mass of $10^{-4}M_\odot$. For $\Delta M=10^{-5}$ and $10^{-3}M_\odot$, PSR J1119-6127 would have TD internal fields. Since the internal field configuration is determined by the sign of $\epsilon_{\rm B}$, to intuitively illustrate the effect of accreted mass on the ellipticity, we show in Fig. \ref{fig:2} the curve of $\left|\epsilon_{\rm B}\right|$ versus $\Delta M$ for PSR J1119-6127. Here we take $t=4.2$ kyr and $\chi=21^\circ$. The blue solid curve represents the $\left|\epsilon_{\rm B}\right|-\Delta M$ evolution when the NS has PD internal fields ($\epsilon_{\rm B}>0$), whereas the red dotted and green dashed curves correspond to the $\left|\epsilon_{\rm B}\right|-\Delta M$ evolutions when the NS has TD internal fields ($\epsilon_{\rm B}<0$). In the insets, the curves of $\epsilon_{\rm B}$ versus $\Delta M$ for the TD case are presented. 

A general evolution trend in Fig. \ref{fig:2} is that $\left|\epsilon_{\rm B}\right|$ increases with the increase of $\Delta M$, and the sign of $\epsilon_{\rm B}$ changes from minus to plus at $\Delta M\simeq5.9\times10^{-5}M_\odot$, and then from plus to minus at $\Delta M\simeq1.1\times10^{-4}M_\odot$.\footnote{Here we stress that neither the increase of $\left|\epsilon_{\rm B}\right|$ nor the changes of NS internal field configuration with the increase of $\Delta M$ is caused by quadrupole deformation from mass accretion since this deformation is neglected in our work for simplicity.} The changes in the sign of $\epsilon_{\rm B}$ correspond to the changes of the internal field configuration. This provides a clear explanation of why the internal field configuration of PSR J1119-6127 varies with $\Delta M$, as presented in the upper panel of Fig. \ref{fig:1}. It should be noted that the discontinuities of $\left|\epsilon_{\rm B}\right|$ are just caused by different expressions of $\epsilon_{\rm B}$ adopted for the PD and TD cases \cite{Cheng:2019}. Fig. \ref{fig:2} shows that $\left|\epsilon_{\rm B}\right|$ could be as small as $\sim10^{-7}$ if $\Delta M=10^{-5}M_\odot$ is assumed. Though the derived ellipticity (after neglecting the sign) of PSR J1119-6127 is larger than the maximum value the NS crust could sustain in general relativity \cite{Gittins:2021}, it is generally consistent with the ellipticity caused by Hall drift of the crustal magnetic fields \cite{Suvorov:2016}. Moreover, the derived $\epsilon_{\rm B}$ of all young pulsars focused in this work are not excluded by the most recent searching results of continuous gravitational waves from isolated pulsars performed by the Advanced LIGO and Advanced Virgo \cite{Abbott:2022} especially when considering that $\Delta M$ may be small. In fact, even if these sources have the maximum ellipticities (when $\Delta M=10^{-3}M_\odot$), their MD radiation still completely dominates over GW radiation.

The effect of the age $t$ of PSR J1119-6127 on the $\xi-\chi$ curve can be found from the comparison of the upper and lower panels of Fig. \ref{fig:1}. Changing $t$ from 4.2 to 7.1 kyr, the resultant $\xi-\-\chi$ curves do not vary when $\Delta M=10^{-5}$ and $10^{-3}M_\odot$ are assumed. However, the $\xi-\-\chi$ curves obtained using the two ages are remarkably different when $10^{-4}M_\odot$ is adopted. Assuming $\Delta M=10^{-4}M_\odot$ and $t=7.1$ kyr, PSR J1119-6127 may have PD if its tilt angle satisfies $\chi\lesssim19^\circ$ and $\chi\gtrsim54^\circ$ (see the thin solid curves in the lower panel). For an intermediate angle $19^\circ\lesssim\chi\lesssim54^\circ$, its internal fields may be TD, as presented by the orange thick solid line. Therefore, the uncertainty in $t$ can affect the diagnosis of the internal field configuration of PSR J1119-6127 when $\Delta M=10^{-4}M_\odot$ is assumed. In summary, in the re-emergence scenario, one cannot simply determine the NS internal field configuration from the observed $n$ as that done in the decay scenario \cite{Hu:2023a,Yan:2024}. Instead, the parameters $\Delta M$, $t$, and $\chi$ all play an important role in determining the internal field configuration of a pulsar, even if it has $n<3$.

The constraints on $\xi$ for PSR J1119-6127 can be obtained from Fig. \ref{fig:1}, in which the vertical black dotted lines from left to right respectively correspond to $\chi=7^\circ$, $16^\circ$, and $21^\circ$ measured (see Tab. \ref{tab:table}). In the upper panel, the black dotted line at $\chi=7^\circ$ from top to bottom intersects with the blue solid, green solid, and red solid lines. The values of $\xi$ at these intersections are respectively $1.13\times10^8$, $1.13\times10^6$, and $5.61\times10^5$. In the lower panel, at $\chi=7^\circ$ three intersections can also be found and $\xi$ at these intersections are respectively $1.13\times10^8$, $1.13\times10^6$, and $6.81\times10^5$. Therefore, given that PSR J1119-6127 may have an tilt angle $\chi=7^\circ$ and an age $t=4.2-7.1$ kyr, constraints on the number of precession cycles are
\begin{eqnarray}
\left\{ \begin{aligned}
         &\xi\simeq1.13\times10^6,~~~~~~~~~~~~~~~~~~~~~~~~~{\rm for}~\Delta M=10^{-5}M_\odot, \\
                &5.61\times10^5\lesssim\xi\lesssim6.81\times10^5,~~~{\rm for}~\Delta M=10^{-4}M_\odot, \\
         &\xi\simeq1.13\times10^8,~~~~~~~~~~~~~~~~~~~~~~~~~{\rm for}~\Delta M=10^{-3}M_\odot.       
                          \end{aligned} \right.
\end{eqnarray}
The results suggest that changing $t$ of PSR J1119-6127 from 4.2 to 7.1 kyr, the constraints on $\xi$ at the angle $\chi=7^\circ$ do not vary when $\Delta M=10^{-5}$ and $10^{-3}M_\odot$ are assumed. However, for $\Delta M=10^{-4}M_\odot$, the value of $\xi$ increases with the increase of $t$, distributing in the range $5.61\times10^5\lesssim\xi\lesssim6.81\times10^5$. For the other two measured angles $\chi=16^\circ$ and $21^\circ$, we perform similar analyses and set constraints on $\xi$ at the two angles. The constraints on $\xi$ of PSR J1119-6127 are presented in Tab. \ref{tab:table2}. We stress that the constraints on $\xi$ are obtained by assuming three typical $\Delta M$, however, without considering other observational facts, e.g., the magnetar-like bursts and surface thermal emissions. After taking into account these observations, the constraints on $\xi$ would be more tight. Compared to the results obtained in the decay scenario \cite{Hu:2023a}, the values of $\xi$ derived in the re-emergence scenario are remarkably larger when $\Delta M=10^{-4}$ and $10^{-3}M_\odot$ are adopted. For a small mass $\Delta M=10^{-5}M_\odot$, the values of $\xi$ obtained in this work are approximately equal to that derived in \cite{Hu:2023a}. In the TD cases, to see how much the internal toroidal field strength can influence the constraints on $\xi$, we also adopt $\bar{B}_{\rm in}=2B_{\rm d,i}$ \cite{Ciolfi:2009} in the calculations of PSR J1119-6127. For comparison, the results are shown by dashed lines in Fig. \ref{fig:1}. Obviously, changing the ratio $\bar{B}_{\rm in}/B_{\rm d,i}$ from 10 to 2 would lower $\xi$ by 5 times since the second term of Eq. (\ref{chidot}) indicates that $\epsilon_{\rm B}$ is proportional to $\xi$.

As mentioned at the beginning of Sec. \ref{sec:results}, besides the constraints on $\xi$, we can also obtain $B_{\rm d,i}$ of PSR J1119-6127 at different $\chi$ measured and $\Delta M$ assumed. The constraints on $B_{\rm d,i}$ are presented in Tab. \ref{tab:table2}. The same as the constraints on $\xi$, the uncertainty in $t$ of PSR J1119-6127 has no effect on the resultant $B_{\rm d,i}$ when $\Delta M=10^{-5}$ and $10^{-3}M_\odot$ are assumed. However, for $\Delta M=10^{-4}M_\odot$, a larger $t$ would lead to a weaker $B_{\rm d,i}$. For instance, at $\chi=7^\circ$, the value $5.25\times 10^{14}$ in Tab. \ref{tab:table2} corresponds to $t=4.2$ kyr, while $3.39\times 10^{14}$ corresponds to $t=7.1$ kyr. Nonetheless, the range of $B_{\rm d,i}$ in this case is \textbf{given as [$3.39\times 10^{14}$, $5.25\times 10^{14}$]} traditionally. The results in Tab. \ref{tab:table2} also show that different $\chi$ only lead to slight differences in $B_{\rm d,i}$, however, relatively large differences in $\xi$ for a specific $\Delta M$. In contrast, for a specific $\chi$ measured, the pulsar's $B_{\rm d,i}$ increases remarkably with the increase of $\Delta M$. PSR J1119-6127 may have an initial dipole field of $\sim8\times10^{13}$ G if a small accreted mass $\Delta M=10^{-5}M_\odot$ is assumed. It may have an even larger initial field of $\sim8\times10^{15}$ G when $\Delta M=10^{-3}M_\odot$ is adopted. Therefore, our results suggest that $B_{\rm d,i}$ of PSR J1119-6127 may be quite strong, reaching the magnetar strength. The strong initial field that submerged into the NS interior is possibly the cause of the burst activities observed in this pulsar \cite{Perna:2011,Göğüş:2016}. From the results of PSR J1119-6127, we can see that the parameters $\Delta M$, $t$, and $\chi$ could affect the estimate of the internal field configuration, $\xi$, and $B_{\rm d,i}$ of this pulsar. Though $\chi$ of some young pulsars can be roughly measured, $\Delta M$ and $t$ are generally hard to determine. As a result, without other valuable observables, we cannot accurately determine the internal field configuration, $\xi$, and $B_{\rm d,i}$ of PSR J1119-6127 in the re-emergence scenario.

The same as PSR J1119-6127, $\xi$ and $B_{\rm d,i}$ of other young pulsars studied in this work also cannot be accurately determined, except for the Crab pulsar. After considering the uncertainty in $t$ of these pulsars, and assuming three typical $\Delta M$, constraints on their $\xi$ and $B_{\rm d,i}$ at specific $\chi$ measured are obtained. The results are listed in Tab. \ref{tab:table2}. As $\chi$ of PSR J1640-4631 is unavailable currently, its results are not presented. Obviously, $B_{\rm d,i}$ of PSR J1734-3333 may have magnetar strength, which is consistent with results of \cite{Ho:2015}. Re-emergence of the strong $B_{\rm d,i}$ of PSR J1734-3333 may make this pulsar behave as a magnetar. A similar conclusion has also been given in \cite{Espinoza:2022}, which suggested that some observed properties of PSR J1734-3333 indeed resemble that of magnetars and in the future it may evolve into a magnetar \cite{Gao:2017}. PSR J1846-0258 also probably has magnetar-strength $B_{\rm d,i}$. Re-emergence of such a strong initial field may be responsible for the magnetar-like outburst observed in PSR J1846-0258 \cite{Hu:2023b}. Taken as a whole, the results in Tab. \ref{tab:table2} suggest that the largest influence on $B_{\rm d,i}$ of these pulsars actually comes from $\Delta M$ assumed. For all typical $\Delta M$ adopted, only PSRs J1734-3333, J1846-0258, and J1119-6127 probably have magnetar-strength $B_{\rm d,i}$. There is a chance that PSRs J1833-1034, J1513-5908, and the Vela pulsar were also born as magnetar. However, this can be true only if relatively large masses of $10^{-4}-10^{-3}M_\odot$ were accreted after their births. Theoretically, if the magnetar-strength magnetic fields are distributed in the NS crust, their evolution could heat the NSs and result in high surface temperatures. This could be a useful criterion that helps to narrow down the range of our results. The argument is as follows. When $\Delta M=10^{-5}M_\odot$ is adopted, PSRs B0833-45, J1833-1034, J1513-5908 actually do not have magnetar-strength $B_{\rm d,i}$ (see Tab. \ref{tab:table2}). This is consistent with the fact that neither magnetar-like bursts nor significantly high surface temperatures are observed for these sources. In other words, the solutions of $\Delta M=10^{-3}M_\odot$ are probably excluded for PSRs B0833-45 and J1833-1034. Also the solutions of $\Delta M=10^{-4}$ and $10^{-3}M_\odot$ could be excluded for J1513-5908. These greatly narrow down the ranges of $B_{\rm d,i}$ and $\xi$ of the three pulsars. Similarly, though $B_{\rm d,i}$ of PSRs J1734-3333, J1846-0258, and J1119-6127 reach magnetar-strength by assuming $\Delta M=10^{-5}M_\odot$, the strengths are approximately equal to the current strengths of dipole fields inferred from $P$ and $\dot{P}$ of these sources. Comparison of the surface thermal emissions of PSRs J1734-3333, J1846-0258 and J1119-6127, and the magneto-thermal evolution calculations indeed does not support the existence of magnetic fields that are much larger than their current dipole fields \cite{Viganò:2013,Hu:2017}. Consequently, the solutions correspond to $\Delta M=10^{-4}$ and $10^{-3}M_\odot$ in Tab. \ref{tab:table2} for the three pulsars may be excluded, while that correspond to $\Delta M=10^{-5}M_\odot$ are still favored.

Since some of the solutions in Tab. \ref{tab:table2} for the young pulsars are possibly excluded, constraints on $\xi$ for these pulsars could be tight. Regarding $\xi$ of these pulsars, the generic feature is that their values generally distribute in the range $\sim10^4-\-10^6$ with the most probable range being $\sim10^4-\-10^5$. Therefore, the results may indicate moderate mutual frictions between superfluid neutrons and other particles in the NS interior in the re-emergence scenario, rather than weak interactions between superfluid neutrons and lattices in the NS crust govern by phonon excitations \cite{Haskell:2018,Haskell and Sedrakian:2018,Cheng:2019}. Assuming $\Delta M=10^{-5}M_\odot$, the values of $\xi$ obtained are approximately equal to that derived in the decay scenario for the pulsars \cite{Hu:2023a,Yan:2024}. Moreover, we note that the allowed values of $\xi$ of the Vela pulsar are consistent with the results from modeling of the rise processes of the large glitch that occurred in this pulsar \cite{Ashton:2019,Hu:2023a}.

\begin{figure}[H]
\centering
\includegraphics[scale=0.5]{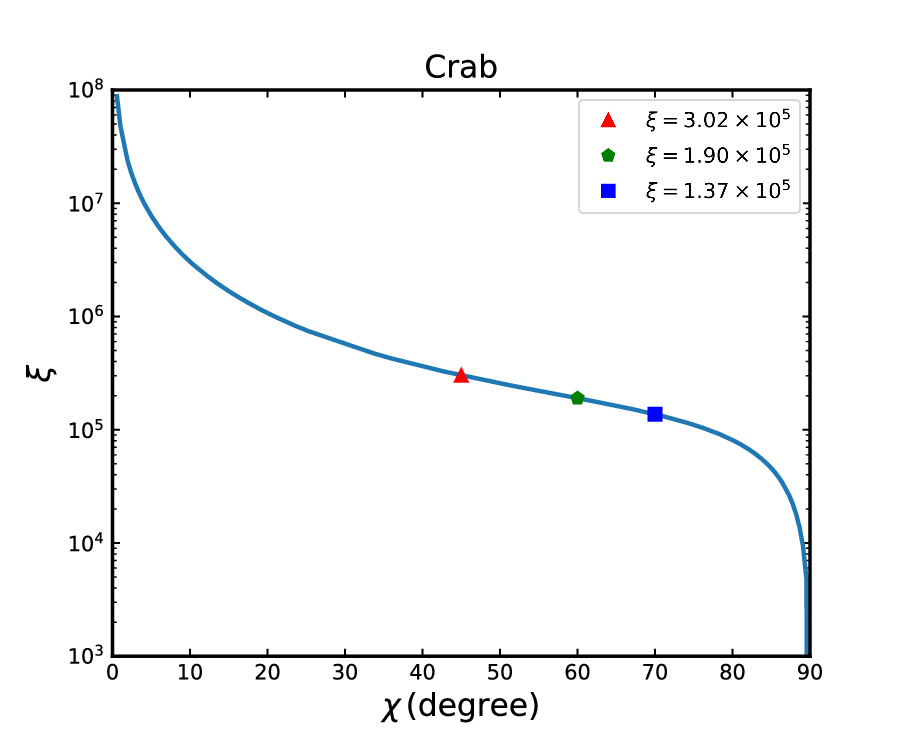}
\caption{$\xi$ versus $\chi$ obtained on the basis of the timing data, age, and inferred $\dot{\chi}$ from observations \cite{Lyne:2013} of the Crab pulsar. The colored triangle, pentagon, and square respectively show the values of $\xi$ at measured $\chi$ of this pulsar (see the legends).}
\label{fig:3}
\end{figure}

Following the same method used in PSR J1119-6127, we also investigate the internal field configuration of PSRs J1734-3333, J1833-1034, J1846-0258, J1513-5908, J1640-4631, and the Vela pulsar. Our detailed calculations show that for their $t$ and $\chi$ listed in Tab. \ref{fig:1}, the young pulsars except for PSR J1640-4631 may have TD internal fields when $\Delta M=10^{-5}$ and $10^{-3}M_\odot$ are adopted. Moreover, PSRs J1734-3333 and J1846-0258, and the Vela pulsar possibly have TD internal fields even if $\Delta M=10^{-4}M_\odot$ is assumed. In contrast, for the same $\Delta M$, the internal fields of PSRs J1833-1034 and J1513-5908 may be PD. Regardless of $\chi$ of PSR J1640-4631, its internal fields are possibly PD for $\Delta M=10^{-4}M_\odot$. However, assuming $\Delta M=10^{-5}$ and $10^{-3}M_\odot$, we could not estimate the internal field configuration of PSR J1640-4631 due to the lack of measured $\chi$. All the results above indicate that we could only obtain relatively rough constraints on the internal field configurations, $B_{\rm d,i}$, and $\xi$ of some young pulsars. As shown below, if their ages and tilt angle change rates could be accurately determined from observations as that of the Crab pulsar, we may set stringent on the internal field configurations, $B_{\rm d,i}$, $\xi$, and $\Delta M$.

\begin{figure}[H]
\centering
\includegraphics[scale=0.5]{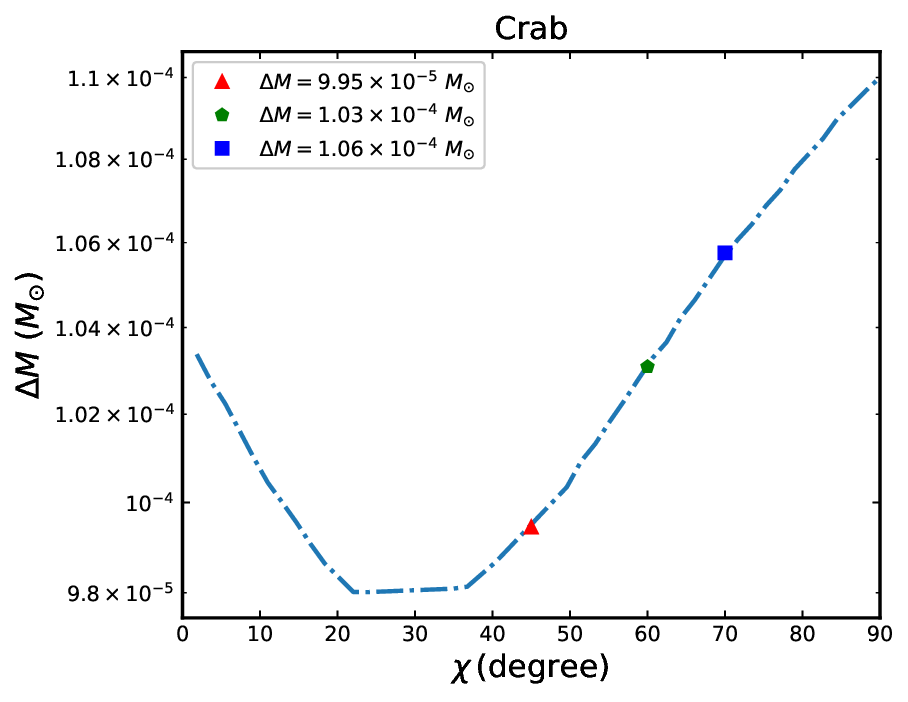}
\caption{$\Delta M$ versus $\chi$ obtained based on the timing data, age, and inferred $\dot{\chi}$ from observations \cite{Lyne:2013} of the Crab pulsar. The colored triangle, pentagon, and square respectively show the values of $\Delta M$ at measured $\chi$ of this pulsar (see the legends).}
\label{fig:4}
\end{figure}

\begin{figure}[H]
\centering
\includegraphics[scale=0.5]{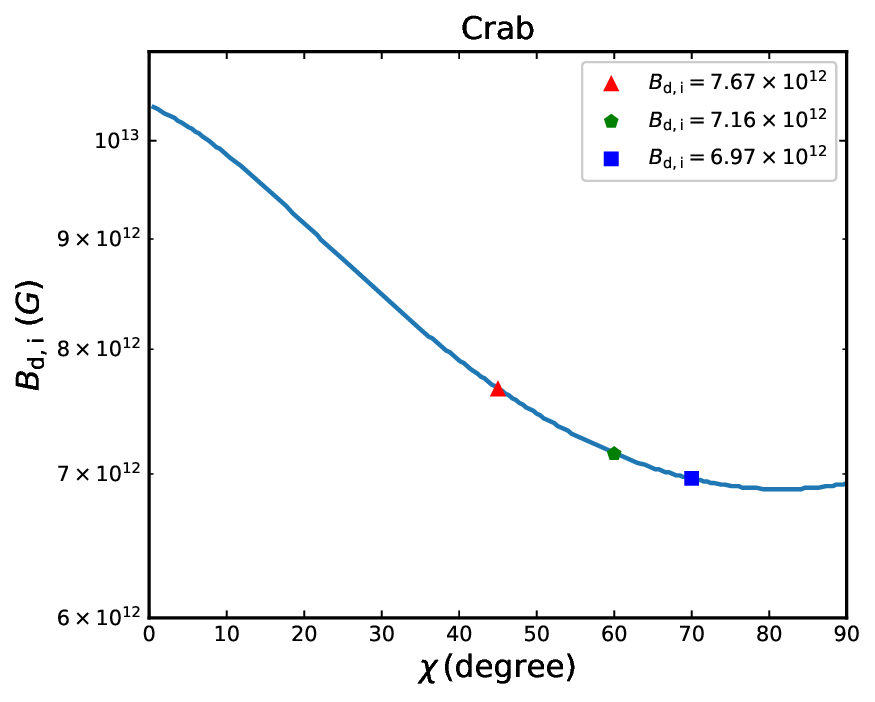}
\caption{$B_{\rm d,i}$ versus $\chi$ obtained based on the timing data, age, and inferred $\dot{\chi}$ from observations \cite{Lyne:2013} of the Crab pulsar. The colored triangle, pentagon, and square respectively show the values of $B_{\rm d,i}$ at measured $\chi$ of this pulsar (see the legends).}
\label{fig:5}
\end{figure}

Lyne et al. \cite{Lyne:2013} reported a steady increase in the separation between the main pulse and interpulse of the Crab pulsar at a rate $0.62^{\circ}\pm0.03^{\circ}$ per century and suggested that a comparable increase rate in the tilt angle may be required to explain the observational results. Here we neglect the error bars for simplicity, the inferred change rate is thus $\dot{\chi}=3.43\times 10^{-12}$ rad/s, which suggests that the internal fields of the Crab pulsar should be TD, as one can see from Eq.  (\ref{chidot}). In previous work, we suggested that to be consistent with the measured timing data, tilt angles, and inferred tilt angle increase rate of the Crab pulsar, depending on the measured $\chi$, its dipole field should increase at a rate $\dot{B}_{\rm d}\sim12-14$ G/s \cite{Yan:2024}. Using the measured timing data, $\chi$, $t$, and inferred $\dot{B}_{\rm d}$, we can set rigorous constraints on its $B_{\rm d,i}$, $\xi$, and $\Delta M$. The curves of $\xi$ versus $\chi$, $\Delta M$ versus $\chi$, and $B_{\rm d,i}$ versus $\chi$ for the Crab pulsar are respectively shown in Figs. \ref{fig:3}, \ref{fig:4}, and \ref{fig:5}. The values of $\Delta M$, $B_{\rm d,i}$, and $\xi$ at measured $\chi$ are presented in Tab. \ref{tab:table2}. In Fig. \ref{fig:3}, we can see that $\xi$ decreases with the increase of $\chi$, and at the measured $\chi=45^\circ$, $60^\circ$, and $70^\circ$ of the Crab pulsar (see Tab. \ref{tab:table}), the number of precession cycles is constrained to be $\xi=3.02\times10^5$, $1.90\times10^5$, and $1.37\times10^5$, respectively. The values of $\xi$ derived here are larger than that obtained in the decay scenario \cite{Hu:2023a}, however, still generally consistent with the results from modeling of the glitch rise processes of some large glitches of the Crab pulsar \cite{Haskell:2018,Hu:2023a}. From Fig. \ref{fig:4}, we find that $\Delta M$ required to produce the inferred $\dot{B}_{\rm d}\sim12-14$ G/s \cite{Yan:2024}, are respectively $9.95\times10^{-5}M_\odot$ for $\chi=45^\circ$, $1.03\times10^{-4}M_\odot$ for $\chi=60^\circ$, and $1.06\times10^{-4}M_\odot$ for $\chi=70^\circ$. Therefore, in the re-emergence scenario, only small $\Delta M$ are needed to account for the Crab's observations. Although the inferred $B_{\rm d,i}$ of the Crab generally show a decreasing trend with the increase of $\chi$, differently measured $\chi$ only result in small differences in $B_{\rm d,i}$ (see Fig. \ref{fig:5}). Specifically, for $\chi=45^\circ$, $60^\circ$, and $70^\circ$, the Crab's $B_{\rm d,i}$ is derived to be $7.67\times10^{12}$, $7.16\times10^{12}$, and $6.97\times10^{12}$ G, respectively. Therefore, current observations of the Crab pulsar suggest that its $B_{\rm d,i}$ may have an ordinary strength of $\sim7\times10^{12}$ G, which is much lower than the typical strength of magnetars. For other young pulsars focused on in this work, to determine their $B_{\rm d,i}$, $\xi$, and $\Delta M$, as well as the internal field configurations, accurate values of $\dot{\chi}$ and $t$ are also needed.

\begin{table*}[]
\caption{\label{tab:table2}%
The tilt angle $\chi$, accreted mass $\Delta M$, initial dipole field $B_{\rm d,i}$, and number of precession cycles $\xi$ for the young pulsars with a steady braking index $n$. For the Crab pulsar, $\Delta M$ at different $\chi$ can be obtained by using $\dot{\chi}$ inferred from observations \cite{Lyne:2013} and the exactly known $t$. For other pulsars, typical values for $\Delta M$ are adopted, as listed in the table. The results for PSR J1640-4631 are not presented due to the lack of measured $\chi$ currently.}
\centering
\tabcolsep 15pt
\renewcommand\arraystretch{1.0}
\begin{tabular}{ccccc}
\toprule
Pulsar& $\chi$& $\Delta{M}~(M_{\odot})$& $B_{\rm d,i}$ (G)& $\xi$\\ \hline
PSR J1734-3333 & $6^{\circ}$ & $10^{-5}$ & $1.02\times 10^{14}$ & $4.16\times 10^5$ \\
{} & {} & $10^{-4}$ & $4.84\times 10^{14}$ & $2.82\times 10^7$ \\
{} & {} & $10^{-3}$ & $1.03\times 10^{16}$ & $4.19\times 10^7$ \\
{} & $21^{\circ}$ & $10^{-5}$ & $9.71\times 10^{13}$ & $2.89\times 10^5$ \\
{} & {} & $10^{-4}$ & $4.33\times 10^{14}$ & $1.15\times 10^7$ \\
{} & {} & $10^{-3}$ & $9.79\times 10^{15}$ & $2.91\times 10^7$ \\
\hline
PSR B0833-45 (Vela) & $62^{\circ}$ & $10^{-5}$ & $4.98\times 10^{12}$ & $5.78\times 10^4$ \\
{} & {} & $10^{-4}$ & [$5.08\times 10^{12}$, $5.27\times 10^{12}$] & [$6.00\times 10^4$, $7.28\times 10^4$] \\
{} & {} & $10^{-3}$ & [$4.77\times 10^{14}$, $4.94\times 10^{14}$] & [$5.83\times 10^6$, $6.05\times 10^6$] \\
{} & $70^{\circ}$ & $10^{-5}$ & $4.85\times 10^{12}$ & $2.90\times 10^4$ \\
{} & {} & $10^{-4}$ & [$4.93\times 10^{12}$, $5.12\times 10^{12}$] & [$3.01\times 10^4$, $3.65\times 10^4$] \\
{} & {} & $10^{-3}$ & [$4.63\times 10^{14}$, $4.80\times 10^{14}$] & [$2.93\times 10^6$, $3.05\times 10^6$] \\
{} & $75^{\circ}$ & $10^{-5}$ & $4.78\times 10^{12}$ & $1.62\times 10^4$ \\
{} & {} & $10^{-4}$ & [$4.87\times 10^{12}$, $5.05\times 10^{12}$] & [$1.68\times 10^4$, $2.04\times 10^4$] \\
{} & {} & $10^{-3}$ & [$4.56\times 10^{14}$, $4.74\times 10^{14}$] & [$1.64\times 10^4$, $1.70\times 10^4$] \\
{} & $79^{\circ}$ & $10^{-5}$ & $4.74\times 10^{12}$ & $8.70\times 10^3$ \\
{} & {} & $10^{-4}$ & [$4.83\times 10^{12}$, $5.01\times 10^{12}$] & [$9.02\times 10^3$, $1.09\times 10^4$] \\
{} & {} & $10^{-3}$ & [$4.52\times 10^{14}$, $4.70\times 10^{14}$] & [$8.81\times 10^5$, $9.15\times 10^5$] \\
\hline
PSR J1833-1034 & $70^{\circ}$ & $10^{-5}$ & $5.14\times 10^{12}$ & $2.64\times 10^4$ \\
{} & {} & $10^{-4}$ & [$7.09\times 10^{12}$, $7.52\times 10^{12}$] & [$4.61\times 10^5$, $6.80\times 10^5$] \\
{} & {} & $10^{-3}$ & $5.18\times 10^{14}$ & $2.73\times 10^6$ \\
\hline
PSR J1846-0258 & $10^{\circ}$ & $10^{-5}$ & $9.48\times 10^{13}$ & $2.93\times 10^5$ \\
{} & {} & $10^{-4}$ & [$8.71\times 10^{14}$, $8.85\times 10^{14}$] & [$3.72\times 10^6$, $3.78\times 10^6$] \\
{} & {} & $10^{-3}$ & $9.52\times 10^{15}$ & $2.91\times 10^7$ \\
\hline
PSR J1119-6127 & $7^{\circ}$ & $10^{-5}$ & $8.03\times 10^{13}$ & $1.13\times 10^6$ \\
{} & {} & $10^{-4}$ & [$3.39\times 10^{14}$, $5.25\times 10^{14}$] & [$5.61\times 10^5$, $6.81\times 10^5$] \\
{} & {} & $10^{-3}$ & $8.09\times 10^{15}$ & $1.13\times 10^8$ \\
{} & $16^{\circ}$ & $10^{-5}$ & $7.80\times 10^{13}$ & $7.67\times 10^5$ \\
{} & {} & $10^{-4}$ & [$3.19\times 10^{14}$, $5.01\times 10^{14}$] & [$3.93\times 10^6$, $1.09\times 10^7$] \\
{} & {} & $10^{-3}$ & $7.86\times 10^{15}$ & $7.69\times 10^{7}$ \\
{} & $21^{\circ}$ & $10^{-5}$ & $7.62\times 10^{13}$ & $5.98\times 10^5$ \\
{} & {} & $10^{-4}$ & [$3.03\times 10^{14}$, $4.82\times 10^{14}$] & [$8.56\times 10^6$, $6.98\times 10^7$] \\
{} & {} & $10^{-3}$ & $7.67\times 10^{15}$ & $6.01\times 10^7$ \\
\hline
PSR J1513-5908 & $3^{\circ}$ & $10^{-5}$ & $3.03\times 10^{13}$ & $2.44\times 10^6$ \\
{} & {} & $10^{-4}$ & $1.75\times 10^{14}$ & $1.70\times 10^4$ \\
{} & {} & $10^{-3}$ & $3.05\times 10^{15}$ & $2.41\times 10^8$ \\
{} & $10^{\circ}$ & $10^{-5}$ & $2.99\times 10^{13}$ & $1.74\times 10^6$ \\
{} & {} & $10^{-4}$ & $1.71\times 10^{14}$ & $1.84\times 10^5$ \\
{} & {} & $10^{-3}$ & $3.01\times 10^{15}$ & $1.73\times 10^8$ \\
\hline
PSR B0531+21 (Crab) & $45^{\circ}$ & $9.95\times 10^{-5}$ & $7.67\times 10^{12}$ & $3.02\times 10^5$ \\
{} & $60^{\circ}$ & $1.03\times 10^{-4}$ & $7.16\times 10^{12}$ & $1.90\times 10^5$ \\
{} & $70^{\circ}$ & $1.06\times 10^{-4}$ & $6.97\times 10^{12}$ & $1.37\times 10^5$ \\
\bottomrule
\end{tabular}
\end{table*}

\section{Conclusion and discussions}\label{sec:conclusion}
The dipole magnetic field of NSs may be submerged into the stellar interior during the fall-back accretion phase soon after the birth of NSs. Subsequently, the buried stronger dipole field may gradually diffuse out to the stellar surface, leading to an increase of the surface dipole field. In the dipole-field re-emergence scenario, using the measured timing data, ages, and tilt angles of several young pulsars with a steady braking index, we have obtained the initial dipole field $B_{\rm d,i}$, and the number of precession cycles $\xi$ of these NSs for typical accreted masses $\Delta M$ assumed. Our results show that in the re-emergence scenario, these NSs have remarkably larger $\xi$ than that derived in the decay scenario if a large accreted mass $\Delta M=10^{-3}M_\odot$ is adopted. The values of $\xi$ in the two scenarios are comparable only when a small accreted mass $\Delta M=10^{-5}M_\odot$ is taken. Moreover, $\xi$ of these pulsars are generally within $\sim10^4-\-10^6$ in the re-emergence scenario. We also find that $B_{\rm d,i}$ of these young pulsars are mainly dependent on $\Delta M$ assumed, and different measured $\chi$ result in relatively small differences in $B_{\rm d,i}$. Among these pulsars, only PSRs J1734-3333, J1846-0258, and J1119-6127 probably have magnetar-strength $B_{\rm d,i}$, which may be comparable to the current dipole fields inferred from their timing data. To sum up, the quantities $\Delta M$, $\chi$, and $t$ could all affect the estimate of $\xi$, $B_{\rm d,i}$, and the internal field configuration of these pulsars. Especially, in the re-emergence scenario, we cannot simply determine the internal field configuration of NSs from their measured braking indices as that done in the decay scenario \cite{Hu:2023b,Yan:2024}. 

Since the tilt angle change rate $\dot{\chi}$ of the Crab pulsar can be deduced from the steady increase in the separation between the main pulse and interpulse observed \cite{Lyne:2013}, by using the value of $\dot{\chi}$, Yan et al. \cite{Yan:2024} derived the dipole field increase rates $\dot{B}_{\rm d}$ at different $\chi$ measured. Based on the resultant $\dot{B}_{\rm d}$, in the re-emergence scenario, we have set stringent constraints on $B_{\rm d,i}$, $\xi$, and $\Delta M$ of the Crab pulsar in this work. The results are presented in Tab. \ref{tab:table2}, which show that the Crab pulsar may have an ordinary $B_{\rm d,i}$, and accreted a small amount of matter after birth. Moreover, the range of $\xi$ derived here is generally consistent with that obtained from modeling of the rise processes of some large glitches of this pulsar \cite{Haskell:2018}. We thus suggest that if $t$ and $\dot{\chi}$ of the young pulsars could be accurately obtained from observations as that of the Crab pulsar, we may set stringent constraints on $B_{\rm d,i}$, $\xi$, $\Delta M$, and the internal field configurations. Constraints on these physical quantities may be useful for the study of transient emissions that are possibly related to NSs, the origin of strong magnetic fields of NSs, pulsar population synthesis, accretion under extreme conditions, and continuous GW emission from pulsars.

Though we have set constraints on some critical physical quantities of NSs above, some issues still remain to be addressed. First, decay of the dipole field due to the combined effects of Hall drift and Ohmic decay (e.g., \cite{Goldreich:1992,Cheng:2019}) is neglected in our calculations. In fact, to set more stringent constraints on $B_{\rm d,i}$, $\xi$, $\Delta M$, and the internal field configuration of these young pulsars, not only $\dot{\chi}$ and $t$ are required, but also both the re-emergence and decay of the dipole field should be considered. Second, it is possible that the spin evolution of the NS may be affected by a long-standing fall-back disk if fall-back matter can circularize around the NS and form a disk. In this case, the effect of the fall-back disk should also be involved in the calculation of the braking index. Third, the accreted matter may pile up and form accretion mountains on the NS surface, leading to an additional deformation \cite{Sur:2021}. In other words, the ellipticity may be contributed by both magnetic field and accretion-induced deformations and have a form that varies from what we used in this work. The effect of the new form of ellipticity on the final conclusion is quite worthy of investigation. It is also interesting to investigate what the continuous GW signals from pulsars will be after taking into account the constraints on $\xi$ obtained here. In future work, we will attempt to resolve these issues. Finally, it should be pointed out that the Crab and Vela pulsars are precessing in our model, whereas no definite observational \textbf{evidence} for the precession of the two pulsars has been found (e.g. \cite{Jones:2001}). The non-detection of the precession of the Crab and Vela pulsars may be attributed to (i) the presence of timing noise \cite{Link:2001}, and (ii) the small amplitudes and long precession periods \cite{Dall'Osso:2017}. However, as suggested in previous work \cite{Zanazzi:2015,Dall'Osso:2017}, the increase of $\dot{\chi}$ as inferred from the Crab's observations \cite{Lyne:2013} is possibly the evidence of precession of this pulsar.

\Acknowledgements{This work is supported by the National SKA program of China (Grant No. 2020SKA0120300), the National Natural Science Foundation of China (Grant Nos. 12003009, and 12033001), and the CAS “Light of West China” Program (Grant No. 2019-XBQNXZ-B-016).}

\InterestConflict{The authors declare that they have no conflict of interest.}

\end{multicols}
\end{document}